\documentclass[twocolumn,showpacs,aps,prl,superscriptaddress,preprintnumbers,amsmath,amssymb,floatfix]{revtex4}

\usepackage{graphicx}    
\usepackage{dcolumn}     
\usepackage{bm}          

\input babarsym

\long\def\inst#1{\par\nobreak\kern 4pt\nobreak
    {\it #1}\par\vskip 10pt plus 3pt minus 3pt}

\RequirePackage{xspace}

\usepackage{relsize}

\newcommand{\GeVc}{\ensuremath{{\mathrm{\,Ge\kern -0.1em V\!/}c}}\xspace}
\newcommand{\MeVc}{\ensuremath{{\mathrm{\,Me\kern -0.1em V\!/}c}}\xspace}
\newcommand{\GeVcd}{\ensuremath{{\mathrm{\,Ge\kern -0.1em V\!/}c^2}}\xspace}
\newcommand{\MeVcd}{\ensuremath{{\mathrm{\,Me\kern -0.1em V\!/}c^2}}\xspace}

\begin{document}

\begin{flushleft}
{\babar-PUB-08/014 \\
SLAC-PUB-13294    } \\
\end{flushleft}

\title{Study of the decay
{\boldmath{$\Ds \rightarrow K^+K^- \ep \nue$}}}

%
\author{B.~Aubert}
\author{M.~Bona}
\author{Y.~Karyotakis}
\author{J.~P.~Lees}
\author{V.~Poireau}
\author{E.~Prencipe}
\author{X.~Prudent}
\author{V.~Tisserand}
\affiliation{Laboratoire de Physique des Particules, IN2P3/CNRS et Universit\'e de Savoie, F-74941 Annecy-Le-Vieux, France }
\author{J.~Garra~Tico}
\author{E.~Grauges}
\affiliation{Universitat de Barcelona, Facultat de Fisica, Departament ECM, E-08028 Barcelona, Spain }
% ...
\collaboration{The \babar\ Collaboration}
\noaffiliation

\date{\today} 

\begin{abstract}
Using 214 $\ensuremath{\mbox{\,fb}^{-1}}\xspace$ of data recorded by the \babar\ detector
at the PEPII electron-positron collider, we study the
decay $D_s^+ \rightarrow K^+K^- e^+ \nu_e$. 
Except for a small S-wave contribution, 
 the events with $K^+ K^-$ masses in the range 1.01-1.03 $\ensuremath{{\mathrm{\,Ge\kern -0.1em V\!/}c^2}}\xspace$ correspond to $\phi$ mesons.
For $D_s^+ \rightarrow \phi e^+ \nu_e$ decays, we measure the relative 
normalization  
of the Lorentz invariant form factors at $q^2=0$, 
$r_V=V(0)/A_1(0)=1.849 \pm0.060 \pm0.095,~r_2=A_2(0)/A_1(0)=0.763 \pm0.071 \pm0.065$ and the pole mass of the axial-vector 
form factors $m_A=(2.28^{+0.23}_{-0.18}\pm0.18)~\ensuremath{{\mathrm{\,Ge\kern -0.1em V\!/}c^2}}\xspace$.
Within the same $K^+K^-$ mass range, we also measure the relative  branching fraction ${\cal B}(D_s^+ \rightarrow K^+ K^- e^+ \nu_e)/{\cal B}(D_s^+ \rightarrow K^+ K^- \pi^+)=0.558 \pm 0.007 \pm 0.016$,  from which we obtain the 
 total branching fraction ${\cal B}(D_s^+ \rightarrow \phi e^+ \nu_e) = (2.61 \pm 0.03 \pm 0.08 \pm 0.15)\times 10^{-2}$.
By comparing this value with the predicted
decay rate, we extract 
$A_1(0) = 0.607 \pm 0.011 \pm 0.019 \pm 0.018$.
The stated uncertainties are statistical, systematic, and  from external inputs.

\end{abstract}


\maketitle

Charm semileptonic decays can help to validate predictions from lattice QCD through precise measurements
of hadronic form factors. 
Such measurements have been 
performed by \babar\ for the $D^0 \rightarrow K^- e^+\nu_e$  decays \cite{kenu}. 
The  $\Ds \rightarrow \phi \ep \nue$ \cite{ref:foot1} channel is 
well suited to study   
form factors in semileptonic decays of charm mesons to a vector particle
because the $\phi$ meson is a narrow resonance which can be well isolated experimentally. 
Because of the higher mass of the spectator $s$-quark,
form factor determinations for this process by lattice QCD are expected to be 
more accurate than for non-strange $D$ mesons.
However, measurements of this decay mode are impacted by the lower production rate 
for $D_s^+$ mesons and higher backgrounds. 
Form factors in $D_s^+ \rightarrow \phi e^+ \nu_e$
 have been previously studied by  photoproduction experiments,
at Fermilab \cite{ref:e653phi,ref:e687phi,ref:e791phi,ref:focusphi}, and
by CLEOII at the CESR $e^+e^-$ collider also operating at the $\Upsilon(4S)$ \cite{ref:cleophi}.
In charm meson semileptonic decays, a $\phi$ meson is expected to originate
only from the $D_s^+$. A possible contribution from 
the Cabibbo suppressed $D^+ \rightarrow \phi e^+ \nu_e$ decay, through the 
$d\bar d$ component of the $\phi$ meson \cite{pdg06} is neglected 
\cite{ref:foot12}.

Using 214 \invfb of data  collected at the $\Upsilon(4S)$ resonance
by the \babar\ detector, we measure 
the $D_s^+\rightarrow K^+K^-e^+\nu_e$ channel decay characteristics,
for events produced in the continuum
$e^+e^- \rightarrow c{\bar c}$.
The analysis focuses on the $\phi e^+ \nu_e$ final state 
in the $K^+K^-$  invariant mass range 
between 1.01 and 1.03 $\GeVcd$. 
The $\phi$ resonance is dominant in this $K^+K^-$ invariant mass region
although a small S-wave component is observed, for the first time, through 
its interference with the $\phi$.

The differential decay rate  for   $D_s^+ \rightarrow K^+ K^- e^+ \nu_e$ 
depends on five variables \cite{ref:cab1}:
$m_{KK}^2$, the mass squared of the $K^+K^-$ system; $q^2$, the
mass squared of the $e^+\nu_e$ system; $\cos{\theta_e}$ 
($\cos{\theta_K}$), where $\theta_e$ ($\theta_K$)
is the angle between the momentum of the $e^+$ ($K^+$) in the $e^+\nu_e$
($K^+K^-$) 
rest frame and the momentum of the $e^+\nu_e$ ($K^+K^-$) system in the 
$D_s^+$ rest frame; 
and $\chi$, the angle between the normals to the planes defined in the 
$D_s^+$ rest frame by the $K^+K^-$ pair and the $e^+\nu_e$ pair.
When analyzing a $D_s^-$ candidate, the 
direction of the $K^-$ is used in place of the $K^+$ 
and $\chi$ is changed to $-\chi$.
The expression for the differential decay rate as a function of these 
variables is given in ref. \cite{ref:wise1}. 
Neglecting contributions proportional to the square of the electron mass,
it depends on three hadronic 
form factors
which are related to the three possible helicity values of
the hadronic current. Restricting to S- and P-wave contributions, 
these form factors  can be written as:
\begin{equation} 
{\cal F}_1 = {\cal F}_{10} + {\cal F}_{11} \cos{\theta_K},~
{\cal F}_2 = \frac{1}{\sqrt{2}}{\cal F}_{21},~
{\cal F}_3 = \frac{1}{\sqrt{2}}{\cal F}_{31}.
\label{eq:fij}
\end{equation}
The form factors ${\cal F}_{ij}$ depend only on $m^2_{KK}$ and $q^2$;
${\cal F}_{10}$ characterizes the S-wave contribution, whereas the
${\cal F}_{i1}$ correspond to the $\phi$ meson:
\begin{equation}
{\cal F}_{i1} = \sqrt{3\pi}qH_i(q^2,m) {\cal A}_{\phi}(m),
\end{equation}
where the $\phi$ meson  decay amplitude ${\cal A}_{\phi}(m)$
is taken to be a relativistic Breit-Wigner 
distribution
with a mass-dependent width including a Blatt-Weisskopf damping factor
\cite{ref:foot2}.
\noindent The form factors $H_{1,2,3}$ 
can be expressed in terms
of the Lorentz invariant form factors $V$ and $A_{1,2}$ \cite{ref:richman},
for which we assume a $q^2$ dependence dominated by a single pole: 
\begin{equation}
V(q^2)=\frac{V(0)}{1-q^2/m_V^2};~A_{1,2}(q^2)=\frac{A_{1,2}(0)}{1-q^2/m_{A}^2}.
\label{eq:ffpole}
\end{equation}
$m_{A}$ and $m_V$ are the pole masses, usually fixed to 
the values of corresponding resonance masses:
$m_{A}=2.5~\GeVcd$$(\simeq m_{D_{s1}})$ and $m_V=2.1~\GeVcd$$(\simeq m_{D_s^*})$. 
At $q^2=0$, the ratios of the form factors $V$ and $A_2$ relative to 
$A_1$ are denoted by $r_V$ and $r_2$, respectively. 
The S-wave contribution is parameterized assuming $f_0$ production:
\begin{equation}
{\cal F}_{10} =  r_0  \frac{p_{KK}~m_{D_s}}{1-\frac{q^2}{m_A^2}} \frac{ m_{f_0}~ g_{\pi} }{m_{f_0}^2-m^2 -i m_{f_0}
\Gamma^0_{f_0}},
\end{equation}
where $r_0$ is a normalization factor  and $p_{KK}$
is the magnitude of the three-momentum of the $K^+K^-$ system in the $D_s^+$
rest frame. 
The values of the $f_0$  parameters 
($m_{f_0},~g_{\pi}, \Gamma^0_{f_0}$) are taken from Ref.
\cite{ref:bes}.

A detailed description of the detector and the algorithms used
for charged and neutral particle reconstruction and identification is provided
elsewhere~\cite{babar_nim}. 
Monte Carlo (MC) samples of $\Upsilon(4S)$ decays,
charm and other light quarks pairs from continuum events are generated 
using a GEANT4 \cite{ref:geant}. Quark fragmentation, in continuum events,
is described using the JETSET package \cite{ref:jetset}.
Signal MC events are generated with seven times the equivalent
statistics of the data, using a simple pole model for the form factors
with $m_{A}=2.5~\GeVcd$ and $m_V=2.1~\GeVcd$.
The simulation of the characteristics of $D_s^+$ production is 
corrected to account for measured differences compared to data. 
Radiative processes are simulated with PHOTOS \cite{ref:photos}.

We reconstruct $\Ds\rightarrow K^+K^- e^+ \nu_e$ decays, 
for $D_s^+$ produced in  $\epem \rightarrow \ccbar$ events. The 
hadronization of the $\ccbar$ system leads to the formation of two jets, 
emitted back-to-back in the center-of-mass (c.m.) frame. 
The analysis method
is similar to the one used for the decay $D^0 \rightarrow K^- e^+ \nu_e$ 
\cite{kenu}.
The only differences are that the cascade from a $D^*$ is not used
to evaluate the signal, and 
the detector performance for the $\Ds$ reconstruction
is measured using  $\Ds \to\phi \pi^+ $  decays
rather than the cascade decay  $D^{*+}\to D^0 \pi^+$, $D^0\to K^-\pi^+$. 

The event thrust axis is determined from all
charged and neutral particles in the c.m. system and its direction is required to be in the
range $|\cos(\theta_{{\rm thrust}})|<0.6$ to minimize the loss of particles 
in regions close to the beam axis. A plane perpendicular to the thrust axis is used to define two 
hemispheres, equivalent to the two jets produced by quark fragmentation. In each hemisphere, we 
search for the decay products of the $\Ds$, namely a positron, of
momentum greater than 0.5 $\GeVc$,
 and 
two oppositely charged kaons. 
Since the $\nu_e$ momentum is unmeasured, a kinematic fit is performed, constraining the invariant mass of the candidate
$K^+ K^- e^+\nu_e$ system to the $\Ds$ mass. In this fit,
the $\Ds$ direction and the neutrino energy are estimated from the other,
charged and neutral, particles 
measured in the event. The $\Ds$ direction is taken as the direction opposite
to the sum of the momenta of all reconstructed particles, 
except for the kaons and the
positron associated with the signal candidate. 
The neutrino energy is estimated as the difference between the total energy of the jet containing the candidate
and the sum of the energies of all reconstructed particles in that hemisphere. 
The $\Ds$ candidate is retained
if the $\chi^2$ probability of the kinematic fit exceeds 10$^{-2}$.

Sizable backgrounds arise from $\Upsilon(4S)\to B\bar{B}$ decays and 
two-jet events from $e^+e^-\to q \bar{q},~q=u,~d,~s,~c$.
Backgrounds are predominantly rejected by using two Fisher 
discriminant variables that exploit differences in the production 
characteristics of hadrons in signal and background. The first variable is used 
to separate signal in jet-like $c\bar{c}$ events from $B\bar{B}$  
with a more spherical topology. The chosen cut retains 71$\%$ of the signal 
and rejects 86$\%$ of the $B\bar{B}$  background. 
The second Fisher discriminant
uses variables related to the different production characteristics of 
particles from $D_s$ decays and $c$-quark fragmentation.  
The selected cut retains 71$\%$ of the signal decays, and rejects 
72$\%$ of the background.
\begin{figure}[!tbp]
\begin{center}
\mbox{
\includegraphics[width=9cm]{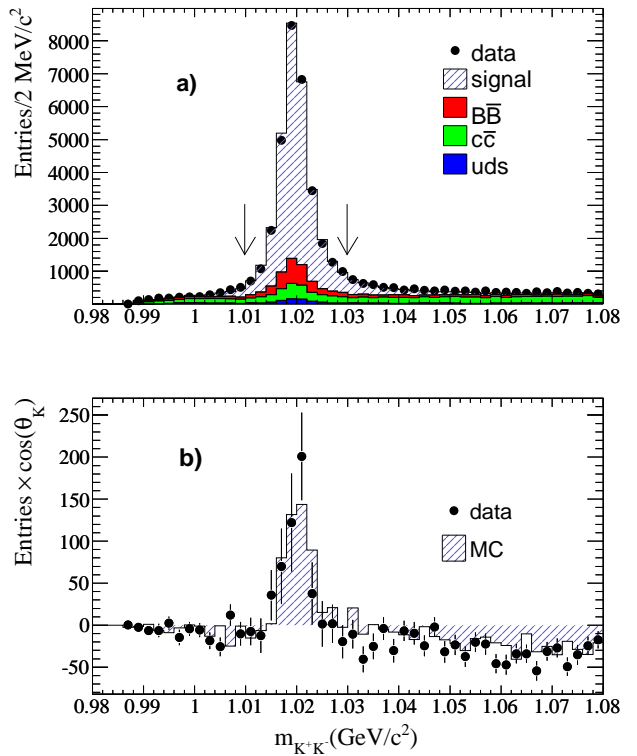}
}
\caption{ a)  $K^+K^-$ invariant mass distribution from data and simulated events. MC events have been normalized to the data luminosity according to the different cross sections. The arrows indicate the selected $K^+K^-$ mass interval. In b), each event is weighted by the measured value
of $\cos{\theta_K}$. Negative entries are produced by the $c{\bar c}$ background asymmetry in $\cos{\theta_K}$.}
\label{fig:mkk}
\end{center}
\end{figure}
The overall signal efficiency 
is approximately 4.5$\%$.
Figure \ref{fig:mkk}a) shows the $K^+K^-$ invariant mass distribution for the selected decays
 compared to the simulation. 
There are 31,839 events in the signal region, with an estimated background of
$20.3\%$.
About $70\%$ of the total background is peaking, corresponding to a $\phi$ decay combined with an electron from another source.
The interference between S- and P- waves generates an asymmetry in the 
$\cos{\theta_K}$ distribution 
which is revealed in Fig. \ref{fig:mkk}b), where  events have been weighted 
by $\cos{\theta_K}$.

To extract $N_S$ (the number of reconstructed signal events),
$r_V$, $r_2$, $m_A$
and $r_0$, we perform a binned maximum likelihood fit to the four-dimensional
 decay distribution in the variables 
$q^2$, 
$\cos\theta_e$, $\cos\theta_K$  and $\chi$. The sensitivity to $m_V$ is  weak and  we fix this parameter to $2.1~\GeVcd$. 
The data are divided into 625 bins, with five equal-sized bins per variable, and
\begin{equation}
{\cal L} = - \sum_{i=1}^{\rm 625} \ln {{\cal P}(n^{\rm data}_i |n^{\rm MC}_i) }.
\end{equation}
\noindent
For each bin $i$, ${\cal P}(n^{\rm data}_i |n^{\rm MC}_i)$ is the
Poisson probability to observe $n^{\rm data}_i$ events when
$n^{\rm MC}_i$ events are expected,
\begin{equation}
n^{\rm MC}_i(\vec{\lambda}) = N_S \frac{\sum_{j=1}^{n_i^{\rm SMC}} w_j(\vec{\lambda})}{W_{\rm tot}(\vec{\lambda})}~+~ n^{\rm BMC}_i.
\end{equation}
\begin{figure}[!tbp]
  \begin{center}  
  \mbox{
\includegraphics[width=9cm]{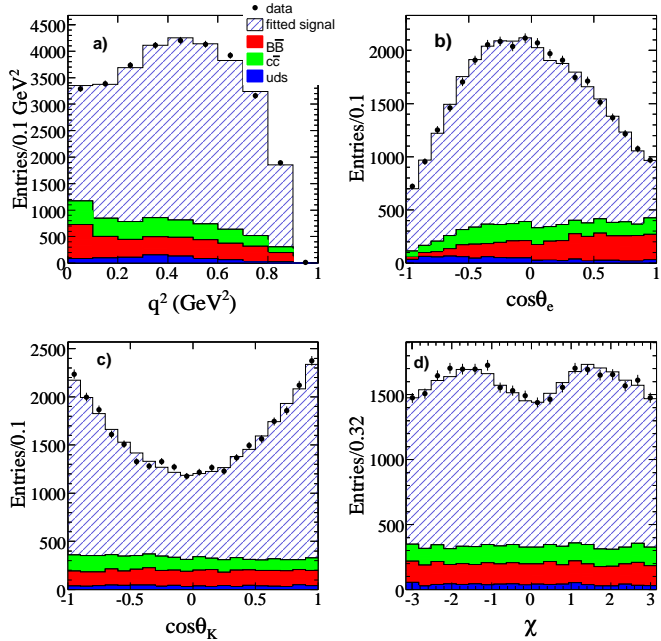}}
 \end{center}
  \caption[]{ Projected distributions of the reconstructed values
of the four kinematic variables.
The data (points with statistical uncertainties) are compared to the sum of four 
histograms, the fitted signal (light hatched) and the estimated background 
contributions (dark shaded).}
   \label{fig:fitdatama}
\end{figure}
Here $n_i^{\rm SMC}$ is the number of signal MC events with reconstructed values of the four variables corresponding to bin $i$
and $n^{\rm BMC}_i$ is the number  of estimated background events. 
They are obtained from MC simulation, 
corrected
for measured differences between data and simulation. 
 Weights, $w_j$, are evaluated for each event, using the generated values of the 
kinematic variables, thus accounting for resolution effects.
$W_{\rm tot}(\vec{\lambda})=\sum_{j=1}^{N^{\rm SMC}} w_j(\vec{\lambda})$
is the sum of the weights for all simulated signal events ($N^{\rm SMC}$)
and $\vec{\lambda}$ corresponds to the parameters to be fitted.
The data and results of the fit are shown in Fig. \ref{fig:fitdatama}
and listed in Table \ref{tab:systematics_ma}.
From the fit we extract a contribution due to S-P wave 
interference.
The value obtained for $r_0$ corresponds to a S-wave fraction  of $(0.22^{+0.12}_{-0.08})\%$ of the decay rate. 

In the fitting procedure, two sources of statistical fluctuations are not 
included. They originate from
the finite sample of simulated signal events and the estimate of the 
average number of background events in each bin.
These effects are evaluated with parameterized simulations and included in 
the systematic uncertainties.
Other systematic effects 
have been assessed to account for the uncertainties in the $c$-quark
hadronization, the background contributions, and the remaining
uncertainties in the simulation of the detector response. They are summarized in Table \ref{tab:systematics_ma}.

Corrections to the simulation of the $c$-quark fragmentation were
performed iteratively, 
comparing variables used in the event selection for samples of
$D_s^+ \to \phi \pi^+$ decays and applying a weight
which depends on the values of these variables. 
We adopt the observed changes in the fit 
parameters for the last step in this iterative process as an estimate of 
the systematic uncertainty. Furthermore we assume a 30$\%$ uncertainty in the simulation of
radiative effects.

\begin{table}[!tb]
  \caption[]{ {Measured values of the parameters  $N_S$, $r_V$, $r_2$, 
$m_A~(\GeVcd)$,
$r_0~({\rm GeV^{-1}})$, their 
statistical and systematic uncertainties.
}
  \label{tab:systematics_ma}}
\begin{center}
  \begin{tabular}{lccccc}
\hline    \hline
  &  $N_S$ &  $r_V$ &   $r_2$ &   $m_A$ & $r_0$\\
\hline 
Fitted value & $25341$&$1.849$ & $0.763$ & $2.28$ & $15.1$\\
\hline
Statistical uncertainty & $178$&$0.060$ & $0.071$ & $^{+0.23}_{-0.18}$ & $2.6$\\
\hline
MC statistics &$81$ &$0.029$ & $0.034$ & $0.09$  & $0.0$ \\
$c\bar{c}$ fragmentation  & $5$&$0.031$ & $0.014$ & $0.07$ & $0.4$\\
Background subtraction  & $480$&$0.081$ & $0.049$ & $0.14$  & $0.6$\\
Detector effects  & $12$&$0.021$ & $0.021$ & $0.07$  & $0.6$\\
    \hline
Total syst. uncertainties&$488$ &$0.095$ & $0.065$ & $0.18$  & $1.0$ \\
\hline \hline
  \end{tabular}
\end{center}
\end{table}

The peaking and combinatorial background components 
from $e^+e^-\rightarrow c\bar{c}$ events
have been studied separately. 
The peaking background contributions are studied by measuring 
inclusive $\phi$ production in events with a fully reconstructed $D^{*+}$
or $D_s^+$ decay. 
The combinatorial background consists mainly of events with a charged 
lepton, one kaon from a 
$D$ decay and a second kaon from fragmentation.
We have measured the rate, momentum and angular distributions of
$K^{\pm}$ accompanying a 
$D^0,~D^{*+}~{\rm or}~D_s^+$ meson in data and 
corrected the corresponding simulation.

\noindent After these corrections, the $K^+K^-$ distribution for selected 
signal events in MC and data agree to within 10$\%$
above 1.03 $\GeVcd$, and this remaining difference is adopted as the 
uncertainty in the normalization of the combinatorial background. 
The $B\bar{B}$ background is obtained from the difference of the  data recorded at the  $\Upsilon(4S)$ resonance and the data recorded 40 $\MeVc$ below. The related systematic uncertainties
are obtained from the statistical accuracy of these measurements and from the uncertainty ($0.25\%$)
between the relative normalization of the two data samples.
Systematic uncertainties also originate from the simulation of the detector 
response. There are small differences in the efficiencies for charged particle reconstruction and electron and kaon identification. They lead to  
data-MC differences in the reconstruction of the $D_s^+$ direction and 
the neutrino energy. They are estimated using $D_s^+ \to \phi \pi^+$
decays.

We measure the $\Ds \rightarrow K^+ K^- e^+ \nu_e$ branching fraction
relative to the decay,
$\Ds \rightarrow  K^+ K^- \pi^+$ for which we adopt the  $K^+K^-$
mass interval, 1.0095-1.0295 $\GeVcd$,
to match the range used by CLEO-c for the $\Ds \rightarrow  K^+ K^- \pi^+$  branching fraction measurement \cite{ref:cleoBR}. 
Specifically, 
we compare the ratio of rates for the two channels
in data and simulated events so that most systematic uncertainties cancel \cite{kenu}.
In the considered mass intervals, we obtain
$R_{Ds} = \frac{{\cal B}(D_s^+ \rightarrow K^+ K^- e^+ \nu_e)}{{\cal B}(D_s^+ \rightarrow K^+ K^- \pi^+)} 
= 0.558 \pm 0.007\pm0.016$. 

Systematic uncertainties are summarized in 
Table \ref{tab:systrate}.
They originate mainly from selection criteria that are not common for the two channels. 
Differences in the impact of the two Fisher discriminants have been 
estimated by varying the selection cuts and differences in particle identification
for electrons and pions are accounted for.
The uncertainty on $N_S$ is taken from the previous fit;
it is dominated by uncertainties in the
background evaluation. 
\begin{table}[!htb]
  \caption[]{ {Summary of the relative systematic 
uncertainties on $R_{Ds}$.}
\label{tab:systrate}}
\begin{center}
  \begin{tabular}{lc}
\hline    \hline
Source & Relative variation \\
\hline
Fisher variable against \ccbar events & $\pm 1.77\%$\\
Fisher variable against \bbbar events & $\pm 0.58\%$\\
Fitted signal ($N_S$)& $\pm1.92\%$\\
PID corrections  & $\pm0.74\%$ \\
$D_s^+$ production & $\pm0.20\%$\\
Mass constrained fit& $\pm0.61\%$\\
 \hline
Total systematic uncertainty& $\pm 2.85\%$ \\
\hline \hline
\end{tabular}
\end{center}
\end{table}
We translate the ratio $R_{Ds}$ to a branching fraction, 
using ${\cal B}(D_s^+ \rightarrow  K^+ K^- \pi^+) = (1.99 \pm 0.10 \pm 0.05) \%$ \cite{ref:cleoBR}, correcting for the 
finite mass range used to select signal events 
$(86.37\pm1.22)\%$, subtracting the S-wave contribution,
and taking ${\cal B}(\phi \rightarrow K^+ K^-)~=~(49.2\pm0.6)\%$
\cite{pdg06}. We find:
\begin{equation}
{\cal B}(D_s^+ \rightarrow \phi e^+ \nu_e) = (2.61 \pm 0.03 \pm 0.08 \pm 0.15)\times 10^{-2}, \nonumber
\end{equation}
\noindent where the last quoted uncertainty corresponds to external inputs.

In conclusion, we have studied the decay 
$D_s^+ \to K^+K^- e^+ \nu_e$ with
a sample of approximately 25,000 signal events, which greatly exceeds any previous 
measurement.  This decay is dominated by the $\phi$ vector meson; 
we measure a small S-wave contribution, possibly associated 
with $f_0 \to K^+K^-$, corresponding to $(0.22^{+0.12}_{-0.08}\pm0.03)\%$ of the $K^+K^- e^+\nu_e$ decay rate.  We have extracted form factor parameters 
from a fit to the four-dimensional decay distribution, assuming 
single pole dominance and obtain:
$r_V=V(0)/A_1(0)=1.849 \pm0.060 \pm0.095,~r_2=A_2(0)/A_1(0)=0.763 \pm0.071 \pm0.065$ and the pole mass of the axial-vector 
form factors $m_A=(2.28^{+0.23}_{-0.18}\pm0.18)~\GeVcd$.
For comparison with previous measurements we also 
perform
the fit to the data  with fixed pole masses $m_A=2.5~\GeVcd$  and 
$m_V=2.1~\GeVcd$, ignoring also the small S-wave contribution.
\begin{table}[htbp]
\begin{center}
 \caption[]{ Results from previous experiments and present measurements.
  \label{tab:meast}}
  \begin{tabular}{lcc}
    \hline \hline
Experiment & $r_V$ & $r_2$  \\
\hline
 E653 \cite{ref:e653phi}  &$2.3^{+1.1}_{-0.9}\pm0.4$ & $2.1^{+0.6}_{-0.5}\pm0.2$  \\
\hline
 E687 \cite{ref:e687phi}  &$1.8 \pm0.9 \pm0.2$ & $1.1 \pm0.8 \pm0.1$  \\
\hline
 E791 \cite{ref:e791phi}  &$2.27 \pm0.35\pm0.22$ & $1.57 \pm0.25\pm0.19$  \\
\hline
 FOCUS \cite{ref:focusphi}   &$1.549 \pm 0.250\pm0.145$ & $0.713 \pm0.202\pm0.266$  \\
\hline
 CLEOII \cite{ref:cleophi}  &$0.9 \pm0.6 \pm0.3$ & $1.4 \pm0.5 \pm0.3$  \\
\hline
 $\babar$ &$1.807\pm0.046\pm0.065$ & $0.816\pm0.036\pm0.030$ \\ 
\hline \hline
  \end{tabular}
\end{center}
\end{table}
The results on $r_2$ and $r_V$ represent a large improvement in statistical 
and systematic precision compared to
earlier measurements  ~\cite{ref:e653phi,ref:e687phi,ref:e791phi,ref:focusphi,ref:cleophi} (see  Table \ref{tab:meast}).

We also  measure the relative  branching fraction ${\cal B}(D_s^+ \rightarrow K^+ K^- e^+ \nu_e)/{\cal B}(D_s^+ \rightarrow K^+ K^- \pi^+)=0.558 \pm 0.007 \pm 0.016$, from which we obtain the 
 total branching fraction ${\cal B}(D_s^+ \rightarrow \phi e^+ \nu_e) = (2.61 \pm 0.03 \pm 0.08 \pm 0.15)\times 10^{-2}$. By comparing this quantity
with the predicted
decay rate, using the fitted parameters for the form factor pole ansatz
we extract 
$A_1(0) = 0.607 \pm 0.011 \pm 0.019 \pm 0.018$. 
Here the third uncertainty refers to the combined value from external 
inputs, namely 
the branching fractions of the $D_s^+$ into $K^+K^-\pi^+$ and
of the $\phi$ into $K^+K^-$, the $D_s^+$ lifetime 
$[(500 \pm7) \times 10^{-15}\mathrm{s}]$ and $V_{cs}=0.9729 \pm 0.0003$. 
Predictions for this decay channel of lattice QCD  calculations,
in the quenched approximation \cite{ref:lqcdphi}, give:
$r_V=1.35^{+ 0.08}_{-0.06}$, $r_2=0.98 \pm0.09$, $m_A=2.42^{+0.22}_{-0.16}~\GeVcd$ and $A_1(0) = 0.63 \pm 0.02$.
They agree with our determination of $A_1(0)$, $r_2$ and $m_A$, but are lower than the measured value 
of $r_V$. 
The measured form factor's ratio $r_2$ is in agreement with 
the value obtained for the same parameterization for the vector decay 
$D \to \bar{K}^* e^+ \nu_e$, whereas $r_V$ is two standard deviations
higher \cite{pdg06}.  
The branching fraction presented here agrees well with the value
($2.68\pm0.13)~\%$, consistent with the assumption of equal semileptonic  decay
widths for the different charm mesons.

\begin{acknowledgments}
We are grateful for the excellent luminosity and machine conditions
provided by our \pep2\ colleagues, 
and for the substantial dedicated effort from
the computing organizations that support \babar.
The collaborating institutions wish to thank 
SLAC for its support and kind hospitality. 
This work is supported by
DOE
and NSF (USA),
NSERC (Canada),
CEA and
CNRS-IN2P3
(France),
BMBF and DFG
(Germany),
INFN (Italy),
FOM (The Netherlands),
NFR (Norway),
MES (Russia),
MEC (Spain), and
STFC (United Kingdom). 
Individuals have received support from the
Marie Curie EIF (European Union) and
the A.~P.~Sloan Foundation.

\end{acknowledgments}


\begin{thebibliography}{99}
\bibitem{ref:foot1}
Charge conjugate states are implied throughout this Letter.
\bibitem{ref:foot12}
Using a singlet-octet mixing angle of 39$^\circ$, the branching fraction
${\cal B}(D^+\rightarrow \phi e^+\nu_e)\sim 2 \times 10^{-6}$ and
we expect less than 10 events from this source in our data sample.
\bibitem{ref:foot2}It is normalized such that: 
$\int_{m_{\phi}-\delta}^{m_{\phi}+\delta}(p^*/m) |{\cal A}_{\phi}(m)|^2dm^2\simeq2\arctan(2\delta/\Gamma^0_{\phi})$ where $\delta$ is a small mass interval and 
$\Gamma^0_{\phi}$ is the total width of the $\phi$. The breakup momentum of the
$\phi$ is denoted as $p^*$.
\end{thebibliography}
\end{document}